\documentclass[preprint2]{emulateapj}

\usepackage{graphics,graphicx}
\usepackage{color}
\usepackage{natbib}
\usepackage{epstopdf}

\newcommand{\chandra}{\textit{Chandra}}
\def\arcsec{\hbox{$^{\prime\prime}$}}

\def\pasa{PASA}

\shorttitle{Young Radio Sources}
\shortauthors{Siemiginowska et al.}

\begin{document}

\title{X-ray properties of the Youngest Radio Sources and Their Environments}

\author{Aneta Siemiginowska$^{1}$, Ma\l gosia Sobolewska$^{1,2}$, Giulia Migliori$^3$,
Matteo Guainazzi$^{4,5}$, Martin Hardcastle$^{6}$, Luisa Ostorero$^{7}$, \L ukasz Stawarz$^{8}$}

\affil{$^1$ Harvard Smithsonian Center for Astrophysics, 60 Garden St, Cambridge, MA 02138, USA}
\affil{$^2$ Nicolaus Copernicus Astronomical Center, Bartycka 18, 00-716 Warsaw, Poland}
\affil{$^3$ Lab. AIM/Univ. Paris Diderot Paris 7/CNRS/CEA-Saclay, France}
\affil{$^4$ Institute of Space and Astronautical Science, JAXA, 3-1-1 Yoshinodai, Sagamihara, Kanagawa, 252-5210, Japan}
\affil{$^5$ European Space Astronomy Centre of ESA, PO Box 78, Villanueva de la Cañada, E-28691 Madrid, Spain}
\affil{$^6$ School of Physics, Astronomy and Mathematics, University of Hertfordshire, College Lane, Hatfield AL10 9AB, UK}
\affil{$^7$ Dipartimento di Fisica - Universit\`a degli Studi di Torino and Istituto Nazionale di Fisica Nucleare (INFN), Via P. Giuria 1, 10125 Torino, Italy}
\affil{$^8$ Astronomical Observatory, Jagellonian University, ul. Orla 171, 30-244 Krak\'ow, Poland}

\smallskip
\email{asiemiginowska@cfa.harvard.edu}


\label{firstpage}

\begin{abstract}

  We present the results of the first X-ray study of a sample of 16
  young radio sources classified as Compact Symmetric Objects (CSOs).
  We observed six of them for the first time in X-rays using {\it
    Chandra}, re-observed four with the previous {\it XMM-Newton} or
  {\it Beppo-SAX} data, and included six other with the archival data.
  All the sources are nearby, $z<1$ with the age of their radio
  structures ($<3000$~years) derived from the hotspots advance
  velocity.  Our results show heterogeneous nature of the CSOs
  indicating a complex environment associated with young radio
  sources. The sample covers a range in X-ray luminosity,
  $L_{2-10\,\rm keV} \sim 10^{41}$--$10^{45}$\,erg\,s$^{-1}$, and
  intrinsic absorbing column density of $N_H \simeq
  10^{21}$--10$^{22}$\,cm$^{-2}$.  In particular, we detected extended
  X-ray emission in 1718$-$649; a hard photon index of $\Gamma \simeq
  1$ in 2021$+$614 and 1511$+$0518 consistent with either a Compton
  thick absorber or non-thermal emission from compact radio lobes, and
  in 0710$+$439 an ionized iron emission line at
  $E_{rest}=(6.62\pm0.04)$\,keV and EW $\sim 0.15-$1.4\,keV, and a
  decrease by an order of magnitude in the 2-10 keV flux since the
  2008 {\it XMM-Newton} observation in 1607$+$26.  We conclude that
  our pilot study of CSOs provides a variety of exceptional
  diagnostics and highlights the importance of deep X-ray observations
  of large samples of young sources. This is necessary in order to
  constrain theoretical models for the earliest stage of radio source
  evolution and study the interactions of young radio sources with the
  interstellar environment of their host galaxies.

\end{abstract}


\section{Introduction}
\label{sec:intro}

Theory and simulations predict that radio sources are the most
luminous when they start growing within the very central regions of
their host galaxies \citep{scheuer1974, begelman1989,
  begelman1996,readhead1996a, heinz1998}.  During this initial
expansion a radio source is expected to interact strongly with the
interstellar medium (ISM), inducing shocks, accelerating particles,
heating the ISM and accelerating gas forming an outflow
\citep{begelman1984,begelman1989,bicknell1997,silk1998, reynolds2001}.
Empirical evidence for such interactions comes from observations of
galactic scale X-ray structures in nearby radio galaxies
\citep[e.g.][]{kraft2000, mingo+2011, siemiginowska+2012,
  hardcastle2012}, as well as from correlations between velocities of
optical line-emitting gas and the morphology of radio outflows in
parsec scale radio sources \citep{morganti2003,
  holt2003,morganti2013,tadhunter2014}.  Such interactions, leading to
energy exchange between a radio source and the ISM, are thought to
contribute to feedback that governs the evolution of galaxies over
cosmological time scales \citep[e.g.][]{best2006,croton2006}. At
present, though, we have only been able to study such interactions in
relatively old, $>10^4-10^5$\,yrs, large radio sources, while studies
of young compact systems have been limited.  Although the youngest
sources would only impact the very central regions of their galaxy,
they can strongly influence the sites directly responsible for the AGN
fuel supply, e.g., by heating and accelerating material away from the
central regions \citep[e.g.][]{bicknell1997,holt2008,wagner2011}.
Thus, these youngest sources could be critical to our understanding of
the feedback processes impacting galaxy evolution.


\begin{table*}
{\scriptsize
\caption{\label{tab:cso} CSOs with Known Redshift and Kinematic Age Measurements}
\begin{center}
\begin{tabular}{rlcrrrll}\hline\hline
\# & Source & $z$    & Size & Velocity & Age$*$  & Refs.$^a$   & Refs.$^b$ \\ 
   & name   &        & [pc] & [$c$]    & [yr] & (Age)           & (X-ray observations)  \\
\\ 
\hline
\\
1. & 1718$-$649$^c$         & 0.014 &   2.0 & 0.07 &   91          & (1)  & This work/C\\ 
2. & 1843$+$356             & 0.763 &  22.3 & 0.39 &  180          & (2)  & This work/C\\
3. & 2021$+$614             & 0.227 &  16.1 & 0.14 &  368          & (2)  & This work/C\\
4. & 0035$+$227             & 0.096 &  21.8 & 0.15 &  450          & (1)  & This work/C\\
5. & 0116$+$319$^d$         & 0.059 &  70.1 & 0.45 &  501          & (1)  & This work/C\\
6. & 0710$+$439             & 0.518 &  87.7 & 0.30 &  932          & (2)  & This work/C; (6/N) \\
7. & 1946$+$708             & 0.101 &  39.4 & 0.10 & 1261          & (1)  & This work/C; (7/B) \\
8. & 1943$+$546             & 0.263 & 107.1 & 0.26 & 1308          & (1)  & This work/C\\
9. & 1934$-$638             & 0.183 &  85.1 & 0.17 & 1603          & (1)  & This work/C; (7/B) \\
10.& 1607$+$26$^e$          & 0.473 & 240   & 0.34 & 2200          & (3)  & This work/C; (8/N)\\
11.& 1511$+$0518           & 0.084 &   7.3 & 0.15 &  300          & (4)  & (9/C) \\
12.& 1245$+$676             & 0.107 &   9.6 & 0.16 &  188          & (1)  & (10/N) \\
13.& OQ$+$208$^f$           & 0.077 &   7.0 & 0.10 &  219          & (5)  & (11/N)\\
14.& 0108$+$388             & 0.669 &  22.7 & 0.18 &  404          & (1)  & (6/N)\\
15.& 1031$+$567             & 0.460 & 109.0 & 0.19 & 1836          & (2)  & (6/N)\\
16.& 2352$+$495             & 0.238 & 117.3 & 0.12 & 3003          & (2)  & (6/N)\\
\\
\hline\hline
\end{tabular}
\end{center}
\smallskip

$^a$ (1) Giroletti \& Polatidis (2009); (2) Polatidis \& Conway (2003);
(3) Nagai et al. (2006); (4) An et al. (2012); (5) Luo et al. (2007).\\
$^b$ C: {\it Chandra}; N: {\it XMM-Newton}; B: {\it BeppoSAX};
(6) Vink et al. (2006); (7) Risaliti et al. (2003); (8) Tengstrand et al. (2009);
(9) Kuraszkiewicz et al. (2009); (10) Watson et al. (2009); (11) Guainazzi et al. (2004).\\
$^c$ NGC 6328; $^d$ 4C$+$31.04; $^e$ CTD 093; $^f$ Mkn 668. \\
$^*$
Many different age estimates are given in the literature for each
source, as the measurements depend on radio frequency, radio
components and the number of epochs. Thus, the systematic errors could
be 20\%-40\% and therefore we only use the age in general discussion
and do not perform any statistical correlations with age.
}
\end{table*}

The existing velocity measurements of radio hot spots and spectral
aging studies indicate that Compact Symmetric Objects (CSOs), a
subclass of GigaHertz-Peaked Spectrum (GPS) sources, are among the
{\it  youngest} extragalactic radio sources, with ages between $\sim$100 and
3000\,yrs \citep{owsianik1998,taylor2000,an2012}. Interestingly, there
seems to be an excess of CSOs younger than $\sim$500\,yrs
\citep{gugliucci+2005}. This excess together with a large number of
compact radio sources ($< 1$~kpc) in comparison to large size radio
galaxies \citep{odea1997} may suggest that the radio jet activity is
intermittent \citep{reynolds-begelman1997, czerny+2009, shulevski2012}
or that the compact sources are short-lived \citep{readhead1996b,
kunert2010, kunert2015}.  In either case the youngest CSOs expand
within the central ($ \approx 100$\,pc) regions of the galaxies and contribute
to the feedback.

The young age of the CSO sources has been challenged by a {\it
  frustrated jet} scenario in which the presence of a dense medium
forces the jets to decelerate rapidly and prevents them from
propagating beyond the central parsec scale regions
\citep{vanbreugel1984, gopal-krishna1991, deyoung1993, carvalho1994,
  deyoung1997, perucho2015}.  This would introduce a systematic bias
to the kinematic age measurements.  However, until now there has been
no evidence for a dense enough medium to halt the jets of GPS sources,
and this lack of evidence gives further support to the {\it youth}
scenario (e.g., \citealt{morganti2008}; see however,
\citealt{garcia2007}).  At the same time, recent broad-band low
frequency radio observations reveal inhomogeneous absorber surrounding
compact radio lobes in a few CSOs
\citep[e.g.][]{tingay2015,callingham2015} implying that complex
interactions between expanding jets and the surrounding medium do take
place. X-ray observations of CSOs can potentially resolve this issue
by providing evidence in favor or against the presence of dense
obscuring matter able to frustrate the jets in the youngest sources.

In general the GPS sources are faint in X-rays and studying them in
this waveband became possible only during the last decade, thanks to
the {\it Chandra} X-ray Observatory
\citep[e.g.,][]{siemiginowska+2008,kunert2014} and {\it XMM-Newton}
\citep[e.g., ][]{guainazzi+2006, vink+2006, teng+2009}.  The early
X-ray results indicate that the GPS X-ray emission could be
intrinsically absorbed, although the absorption properties of the GPS
sources are similar to those found in general source populations. The
intrinsic GPS X-ray radiation could be related to the accretion, to
expanding radio lobes, to relativistic jets, or in some cases to the
thermal hot medium found in the central regions of the host galaxy
\citep{guainazzi+2006, stawarz2008, ostorero2010, migliori+2011,
migliori+2012, migliori2014}.

However, only a few CSO sources with known redshift and
measured age has been observed in X-rays
\citep[e.g.,][]{guainazzi+2006,risaliti2003,vink+2006}.  
Thus, we initiated our {\it
  Chandra} program to study the CSO sample defined by requiring the
  availability of kinematic age measurements and $z<1$ (Section~\ref{sec:sample}). Our
  main goal was to establish the X-ray properties of these CSOs,
  determine the origin of their X-ray emission, and study the
  properties of the environment into which they are expanding. We
  describe our methods of X-ray data analysis in
  Section~\ref{sec:xdata}, and present the X-ray analysis results in
  Section~\ref{sec:results}.
We discuss the implications of the CSO
X-ray properties on the models of the high-energy processes
in compact radio sources in Section~\ref{sec:discussion} and conclude
our findings in Section~\ref{sec:conclusions}.  We use the most recent
constraints on the cosmological parameters to convert the observed
fluxes into luminosities (Hinshaw et al. 2013; $\rm H_0 =
69.3$~km\,$s^{-1}$\,Mpc$^{-1}$, $\Omega_m=0.287$ implemented as WMAP9
in the {\tt astropy.cosmology} package, \citet{astropy2013}).

\section{Sample}
\label{sec:sample}

We compiled from the literature a sample of 16 CSO sources with
redshift and expansion velocity of the radio hot spots available
  at the time of our proposed {\it Chandra} project.
Table~\ref{tab:cso} lists these sources and their general radio
properties (the linear size, $LS \simeq 2$--120\,kpc; the expansion
velocity, $v \simeq 0.07$--0.4\,$c$; and the age inferred through the
kinematic argument, $\tau \sim$100--3000\,yrs).

\begin{figure}
\includegraphics[width=7.5cm]{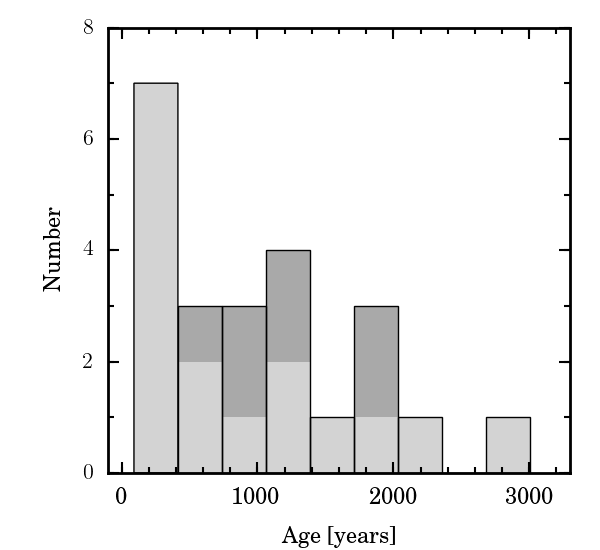}
\caption{Distribution of the CSOs kinematic age for all available
    sources with the age measurements to date. The distribution of
    sources with available X-ray data (Table~\ref{tab:cso}) is marked
    in light grey.  The dark grey indicates additional CSOs from
    \cite{anb2012} (except for PKS~B1413+135 classified as
    core-dominated) without X-ray data, with the ages derived after
    our {\it Chandra} program has been completed. }
\label{fig:ages}
\end{figure}

We obtained new {\it Chandra} observations of 10 CSOs.  We observed
the six CSOs with no prior X-ray information (5\,ksec exposures).
Additionally, we imaged with {\it Chandra} the two sources observed
previously with {\it BeppoSAX} (1946$+$708, 5\,ksec; 1934$-$63,
20\,ksec), or with {\it XMM-Newton} (0710$+$439, 38\,ksec; 1607$+$26,
38\,ksec).  The goal of the long, 20--40\,ksec, exposures was to study
in detail the CSO environments through the high-resolution spatial and
spectral analysis allowed by the {\it Chandra}
data. Table~\ref{tab:chandra} contains the log of our {\it Chandra}
observations.  It includes also 1511$+$0518 with archival 2\,ksec
{\it Chandra} data.

Figure~\ref{fig:ages} shows the distribution of CSOs kinematic ages
for 16 sources in our pilot study
and 7 additional sources from \cite{anb2012} with their ages measured
after our {\it Chandra} program has been completed.  The uncertainties
on the age measurements are typically within $20-40\%$ and related to
the quality of the radio monitoring data. Note that our X-ray sample contains
all the youngest CSOs with the ages $<400$~years.

In summary, Table~\ref{tab:cso} contains a complete list of CSO sources
with derived redshift and kinematic age that have been observed in
X-rays to date, while Table~\ref{tab:chandra} contains the sources
analyzed in this paper. With our {\it Chandra} program, we have
increased the size of the CSO X-ray sample by $\sim$68\%.

\section{{\it Chandra} X-ray Observations and Data Analysis}
\label{sec:xdata}


\begin{table*}
{\scriptsize
\noindent
\caption[]{\label{tab:chandra} {\it Chandra} Observations}
\begin{center}
\begin{tabular}{rlrrrrrrc}
\hline\hline
\# & Source  & RA (J2000) & Dec (J(2000) & Date & Obsid & Exposure & Total$^a$ & Net$^a$ \\
  & name   &            &              &      &  &  [s] & counts  & counts \\
\\
\hline
\\
1. & 1718$-$649    & 17 23 41.0 &	$-$65 00 36.6 & 2010-11-09 & 12849 &  4783 &  231 & 224.9$\pm 15.4$ \\
2. & 1843$+$356    & 18 45 35.1 &	$+$35 41 16.7 & 2010-10-26 & 12850 &  4783 &   11 & 10.8$\pm 3.3$  \\
3. & 2021$+$614    & 20 22 06.7 &	$+$61 36 58.8 & 2011-04-04 & 12853 &  4784 &   54 & 53.8$\pm7.3$ \\
4. & 0035$+$227    & 00 38 08.1 &	$+$23 03 28.4 & 2010-10-13 & 12847 &  4783 &   10 & 9.9$\pm3.2$ \\
5. & 0116$+$319    & 01 19 35.0 &	$+$32 10 50.0 & 2010-11-05 & 12848 &  4742 &    3 & $\dots$ \\ 
6. & 0710$+$439    & 07 13 38.2 &	$+$43 49 17.2 & 2011-01-18 & 12845 & 37845 & 1679 & 1676.8$\pm41.0$\\
7. & 1946$+$708    & 19 45 53.5 &	$+$70 55 48.7 & 2011-02-07 & 12852 &  4742 &  110 & 109.8$\pm 10.5$  \\
8. & 1943$+$546    & 19 44 31.5 &	$+$54 48 07.0 & 2011-05-04 & 12851 &  4783 &   12 & 11.9$\pm 3.5$ \\
9. & 1934$-$63     & 19 39 25.0 &       $-$63 42 45.6 & 2010-07-08 & 11504 & 19793 &  362 & 361$\pm19.0$ \\
10.& 1607$+$26     & 16 09 13.3 &	$+$26 41 29.0 & 2010-12-04 & 12846 & 37845 &  213 & 212.5$\pm 14.5$ \\	
   & secondary$^b$ & 16 09 12.7&       $+$26 41 17.5 & 2010-12-04 & 12846 & 37845 &   32  & 30.8$\pm 5.7$ \\	
11.& 1511$+$0518$^c$ & 15 11 41.2 &      $+$05 18 09.2 & 2003-05-18 & 4047 & 1994 & 49 & 48.8$\pm7.1$\\
\\
\hline\hline
\end{tabular}

\smallskip Notes: $^a$ Total and background subtracted counts in 
 $r = 1.5\arcsec\ $ circle with energies between 0.5\,keV
and 7\,keV; $^b$ the secondary X-ray source resolved in our {\it Chandra} image. $^c$ {\it Chandra} archival data.

\end{center}

}
\end{table*}

The {\it Chandra}\ ACIS-S data were collected during the 2010--2011
epoch (see Table~\ref{tab:chandra} for details).  All targets were
placed at the aim point on the back-illuminated ACIS CCD (S3). The
observations were made in VFAINT mode with 1/8 CCD readout to avoid
pileup if sources were to be bright. All 10 targets were detected by
{\it Chandra}\ with a number of counts between 3 and 1,677.

The X-ray data analysis was performed with the CIAO version 4.6
software \citep{ciao2006} using CALDB version 4.4.  We processed the
data by running CIAO tool {\tt acis-process-events} and applied the
newest calibration files, filtered VFAINT background events, and ran a
sub-pixel event-repositioning algorithm (and set {\tt
  pix\_adj=EDSER}). This final step provides the highest angular
resolution X-ray image data for the most up-to-date ACIS-S
calibration.  All spectral modeling was done in {\it
  Sherpa}\footnote{http://cxc.harvard.edu/sherpa/}
\citep{freeman+2001,refsdal+2009}.  We used the Cash and Cstat fitting
statistics \citep{cash1979} and the Nelder-Mead optimization method
\citep{neldermead1965}.

\subsection{Image Analysis}
\label{sec:images}

 
\begin{figure*}
\includegraphics[width=2.2in]{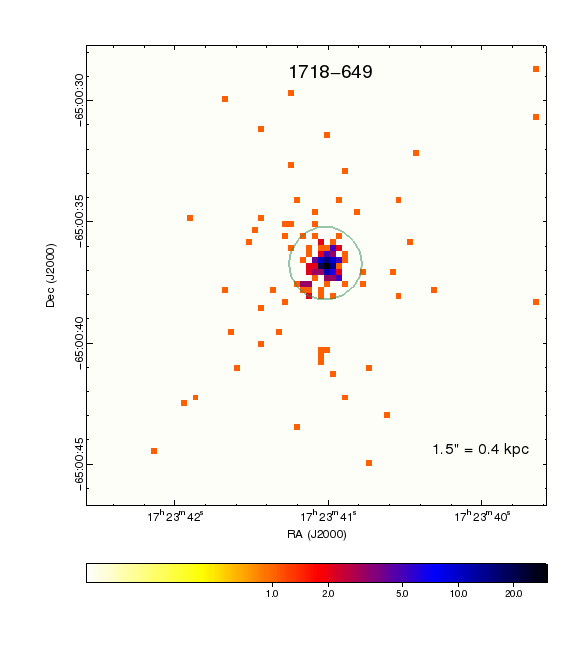}
\includegraphics[width=2.2in]{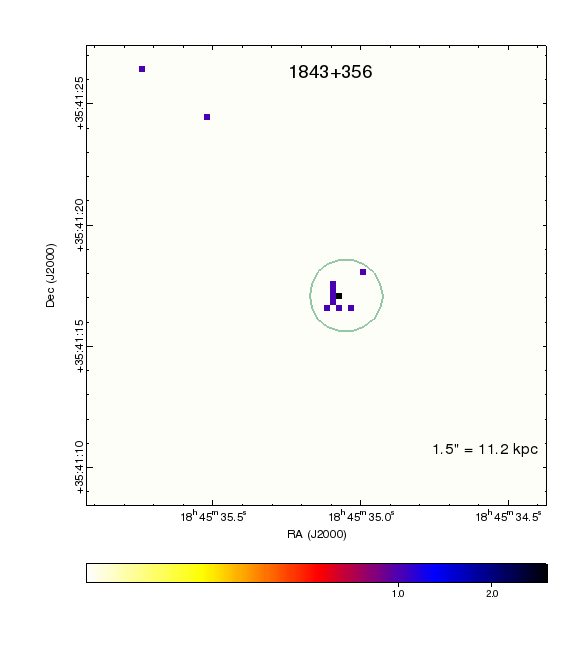}
\includegraphics[width=2.2in]{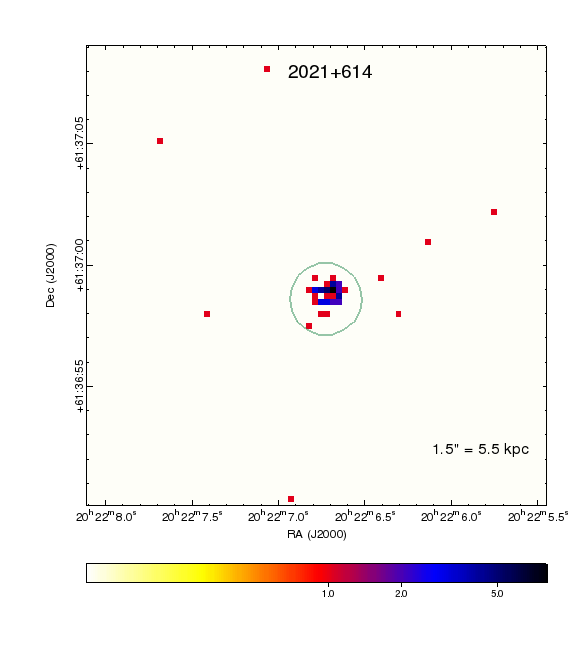}
\includegraphics[width=2.2in]{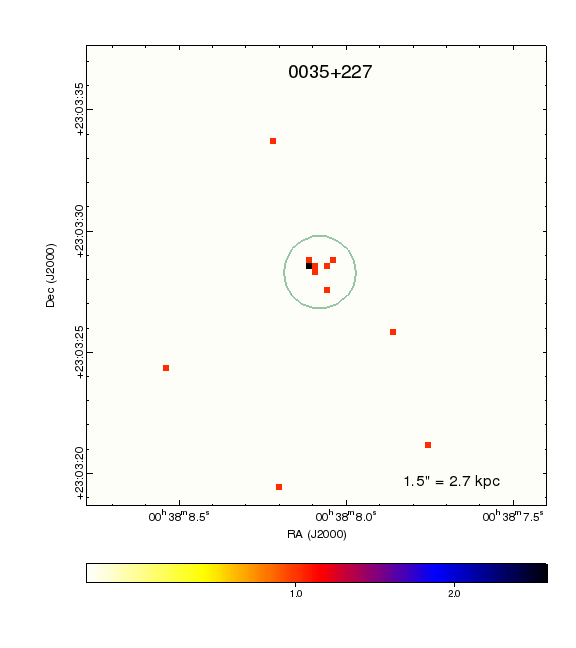}
\includegraphics[width=2.2in]{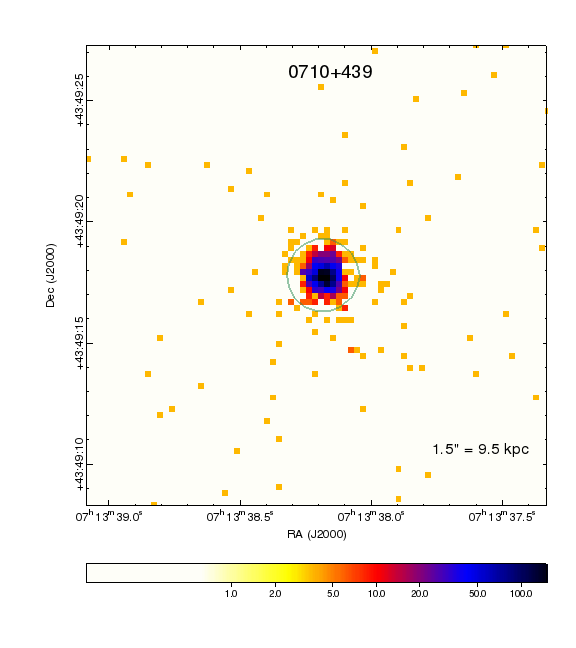}
\includegraphics[width=2.2in]{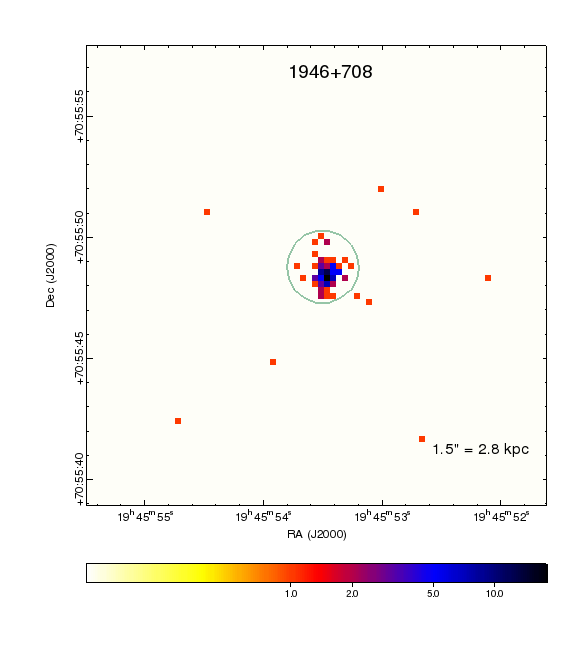}
\includegraphics[width=2.2in]{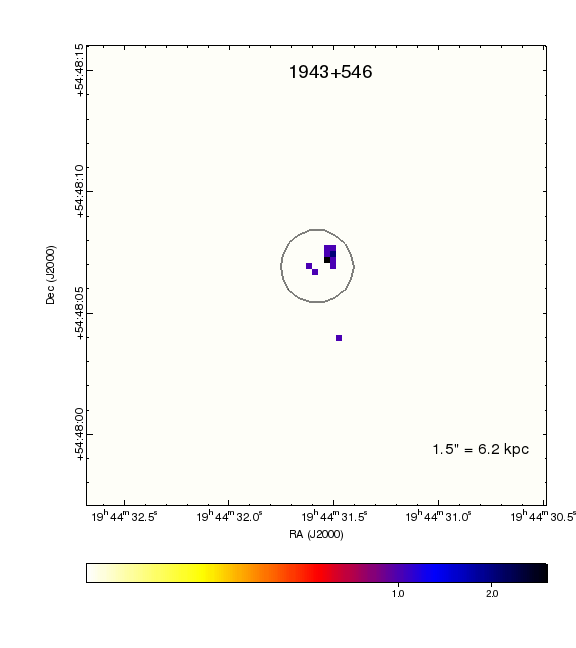}~~~~~~
\includegraphics[width=2.2in]{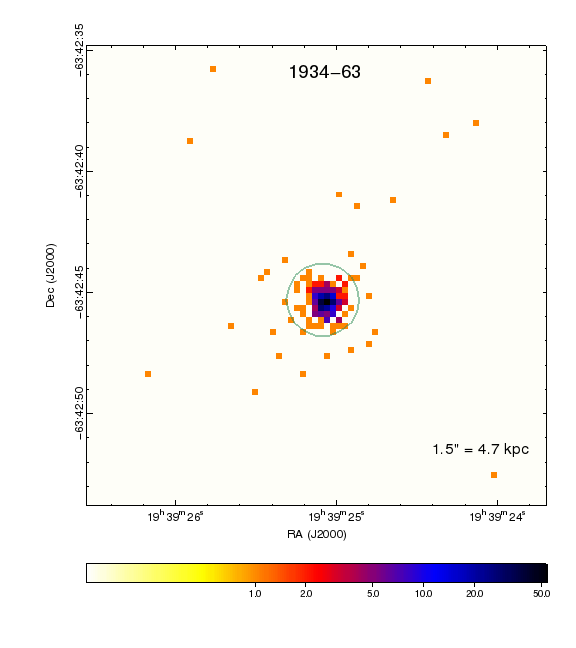}~~~~~~
\includegraphics[width=2.2in]{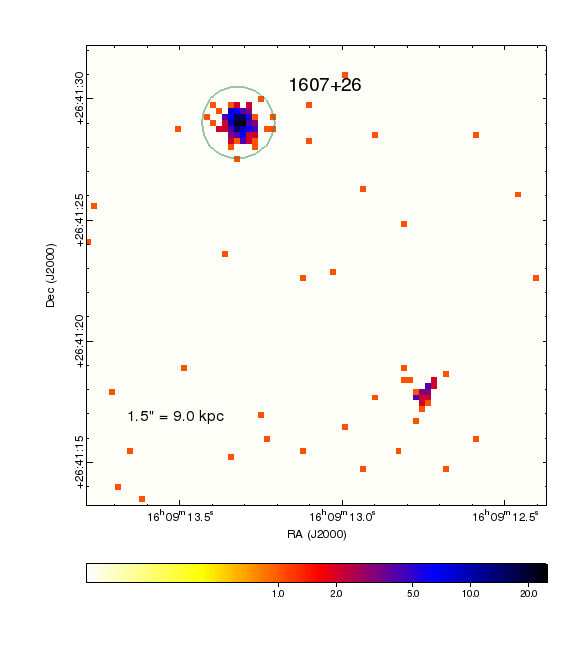}
\caption{ACIS-S images of CSOs sources observed in our {\it Chandra}
  program (Table~\ref{tab:chandra}), except $0116+319$ detected with
  only 3 counts. The X-ray events with the energy range between
  0.5-7~keV are displayed.  The pixel size is set to $0.\arcsec264$
  and each panel shows $9.84\arcsec\times9.84\arcsec$ region. A
  circular region with 1.5\arcsec\, radius is plotted around each source
  and its corresponding scale in kiloparsecs is marked in each panel.
}
\label{fig:images}
\end{figure*}

We inspected the {\it Chandra} data using {\tt ds9} and visually
confirmed the location of each source, and also defined the source and
background regions for further spectral extractions.
Figure~\ref{fig:images} displays the $\sim 10'' \times 10''$ ACIS-S
images of all the targets (denoted in the panels with green circles),
except for $0116+319$ which is detected with 3 counts only ($
p\rm-value \ll 0.01$)\footnote{The $p\rm -value$ is the probability
  that the observed counts are due to the Poisson background
  fluctuations. $ p\rm -value < 0.01$ indicates a detection. It is
  derived via simulations.}.  The target is the brightest source in
the image.  We stress that due to the different redshifts of our CSO
sources, the {\it Chandra} observations probe physical scales that
differ by $\sim$1.5 orders of magnitude (0.4--11.2\,kpc;
Figure~\ref{fig:images}).

Hints of extended X-ray emission in the three sources with the
highest number of counts (0710$+$0439, 1718$-$649, 1934$-$638), or a
secondary X-ray source (in 1607$+$268) could be seen in the images.
We investigated the significance of the extended X-ray emission
using surface brightness analysis. We compared the observed surface
brightness profiles to the {\tt
CHART}\footnote{http://cxc.harvard.edu/chart/} simulated PSF profile
expected for a point source.

\begin{figure*}
\includegraphics[width=\columnwidth]{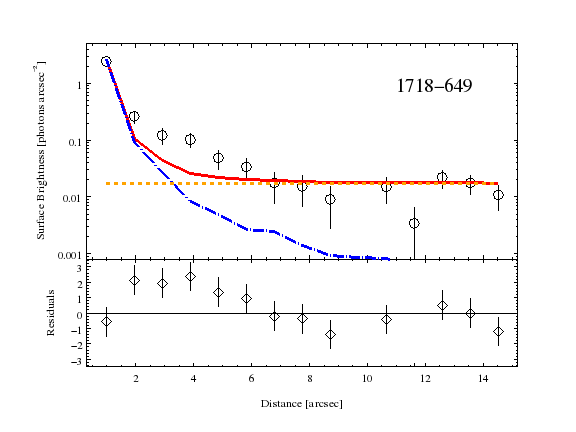}
\includegraphics[width=\columnwidth]{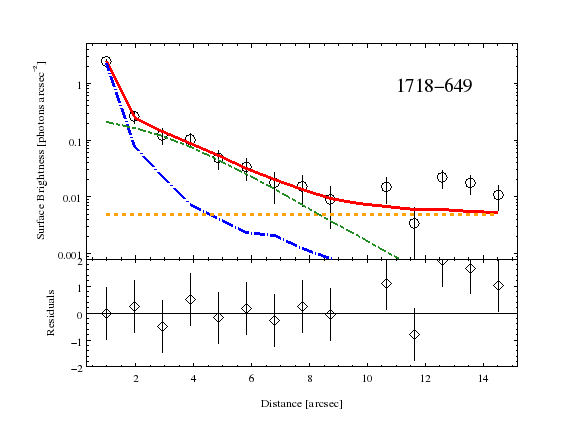}
\caption{X-ray surface brightness profile of 1718$-$649. 
Different model components include the {\it Chandra} PSF profile (blue
long dashed curves), constant background (orange dotted lines), and
the additional standard beta model with $\beta > 0.55$ (green short
dashed curve); thick solid red curves denotes the total model fitted
to the data (without and with the beta component, left and right
panels, respectively), and the corresponding residuals are given in
the lower panels.  }
\label{fig:profile}
\end{figure*}


\begin{table*}
{\scriptsize
\noindent
\caption[]{\label{tab:chandra-fit} {\it Chandra} Best Fit Model Parameters}
\begin{center}
\begin{tabular}{rlccccccr}
\hline\hline
\\
\# & Source  & $N_H^{gal}$            &  $N_H^z$              & $\Gamma$ &  Norm & Flux (soft)   & Flux (hard) & Cstat/d.o.f \\
   & name    & [$10^{20}$\,cm$^{-2}$] &  [$10^{22}$\,cm$^{-2}$] &          &
  [10$^{-6}$\,ph\,cm$^{-2}$\,s$^{-1}$]  &  [$10^{-14}$\,erg\,cm$^{-2}$\,s$^{-1}$] & [$10^{-14}$\,erg\,cm$^{-2}$\,s$^{-1}$] \\
\\
\hline
\\
1. & 1718$-$649   &  7.15  & 0.08$\pm0.07$          & 1.6$\pm0.2$         & 80.2$^{+18.9}_{-15.0}$  & 18.6$\pm3.4$    & 35.3$\pm5.5$  & 334.9/443 \\
2. & 1843$+$356   &  6.75  & 0.8$^{+0.9}_{-0.7}$    & (1.7)               & 4.64$^{+2.1}_{-1.53}$   & 1.1$\pm0.3$   & 2.2$\pm0.9$   & 68.0/444 \\
3. & 2021$+$614   & 14.01  & $<1.02$                & 0.8$^{+0.3}_{-0.2}$ & 10.6$^{+2.7}_{-2.2}$    & 2.7$\pm0.6$     & 19.3$\pm5.6$  &  212.2/443 \\
4. & 0035$+$227   &  3.37  & 1.4$^{+0.8}_{-0.6}$    &  (1.7)              &  8.2$^{+4.1}_{-2.9}$    & 1.8$\pm0.8$     &  3.6$\pm 1.6$ &  72.8/444 \\
5. & 0116$+$319$^a$	   &  5.67  & $\dots$ & $\dots$ & $\dots$ & $<0.5$ & $\dots$ & $\dots$ \\ 
6. & 0710$+$439$^b$ &  8.    & $\dots$ & $\dots$ & $\dots$ & $\dots$ & $\dots$ & $\dots $ \\
7. & 1946$+$708   &  8.57  & 1.7$^{+0.5}_{-0.4}$    & 1.7$\pm0.4$            &  105.3$^{+7.4}_{-4.2}$  & 24.4$\pm12.4$     & 36.6$\pm7.5$  & 296.3/443 \\
8. & 1943$+$546   & 13.15  & 1.1$\pm0.7$            & (1.7)                  & 7.8$\pm2.9$             & 1.8$\pm0.6$   & 3.7$\pm1.4$   & 76.5/444 \\
9. & 1934$-$63    &  6.16  & 0.08$^{+0.07}_{-0.06}$ & 1.67$^{+0.15}_{-0.16}$ & 29.6$^{+4.7}_{-4.1}$    & 5.0$\pm0.5$     & 12.3$\pm1.6$  & 330.2/443  \\
10.& 1607$+$26	  &  4.1   &  $ <0.18$              & 1.4$\pm0.1$            & 7.03$^{+0.85}_{-0.47}$  &  1.6$\pm 0.2$ & 4.8$\pm 1.1$  & 347.1/443 \\	
   & secondary$^c$&  4.1   &  $<0.10$               & 1.4$\pm0.3$            & 1.06$^{+0.35}_{-0.28}$   &  0.2$\pm0.1$  & 0.6$\pm 0.4$ & 143.6/443 \\	
11.& 1511$+$0518   &  3.29  &  $< 0.23$  & 1.0$\pm0.2$    &   22.8$^{+4.8}_{-4.2}$   &  5.6$\pm1.2$     &  31$^{+19}_{-12}$ &  203.8/443 \\
\\
\hline\hline
\end{tabular}

\smallskip Notes: Unabsorbed flux in the observed energy range (soft: 0.5--2\,keV; hard: 2--10\,keV, extrapolated
from the model fitted over the 0.5--7\,keV range);
All uncertainties are given as $1\sigma$ for one interesting parameter; Upper limits are 3$\sigma$ limits;
$^a$ 3 counts were detected in the source region and no model fitting was performed for this source; We list the 3$\sigma$ flux limit estimated using {\tt srcflux} tool which performs simulations \citep{kashyap2010}. 
$^b$ see Table~\ref{tab:0710};
$^c$ the secondary X-ray source resolved in our {\it Chandra} image with $N_H$ limit at z=0.

\end{center}

}
\end{table*}

\subsection{Spectral Analysis}

The X-ray spectra and corresponding calibration files ({\tt arf} and
{\tt rmf}) were extracted with the CIAO script {\sc specextract} for
all the sources.  We assumed a circular region with 1.5$\arcsec$
radius centered on the position of each source for the spectral
extraction. A local background was extracted assuming an annulus with
inner and outer radii equal to $1.7\arcsec$ and $10 \arcsec$,
respectively. We compared the total number of counts and net counts
for each source (Table~\ref{tab:chandra}).  We assessed that the
background was negligible in all cases ($<1$ count in the source
extraction region for all targets, except for 0710$+$439 with 2.1 counts,
e.g.  $<0.1\%$ of the source counts), and therefore we ignored it
during the spectral analysis (i.e. we modeled only the source
spectra).

We applied an absorbed power law model to the spectral data: $N(E) = A
E^{-\Gamma} \rm exp\{-N_H^{gal}\sigma(E) -N_H^{z}
\sigma[E(1+z)]\}$ \,ph\,cm$^{-2}$\,s$^{-1}$\,keV$^{-1}$,
where $A$ is the normalization at 1\,keV, $\Gamma$ is the photon index
of the power law, and $\rm N_H^{gal}$ and $\rm N_H^{z}$ are the
equivalent column densities for the two absorption components,
Galactic and intrinsic to the source.  In the {\it Sherpa}
nomenclature, this baseline spectral model was defined as {\tt
xsphabs*xszphabs*powlaw1d} and fit to the data in the 0.5--7\,keV
energy range. The Galactic absorption was always fixed at the
respective value, $N_H^{gal}$ listed for each source in
Table~\ref{tab:chandra-fit}, while the intrinsic absorption column was
allowed to vary.  The photon index $\Gamma$ was fixed to 1.7 (a
typical value for unabsorbed $N_H<10^{22}$\,cm$^{-2}$ AGN, Burlon et
al. 2011) while modeling the sources with a low number of
counts, $<12$, in order to obtain a limiting value or detection of the
intrinsic absorption. The range of the photon index $1.4 < \Gamma <
2.0$ results in the intrinsic column density consistent to within
1$\sigma$ with the values listed in Table~\ref{tab:chandra-fit}.  The
photon index was a free parameter in sources with a larger number of
counts.


\begin{center}
\begin{figure*}
\includegraphics[width=15cm]{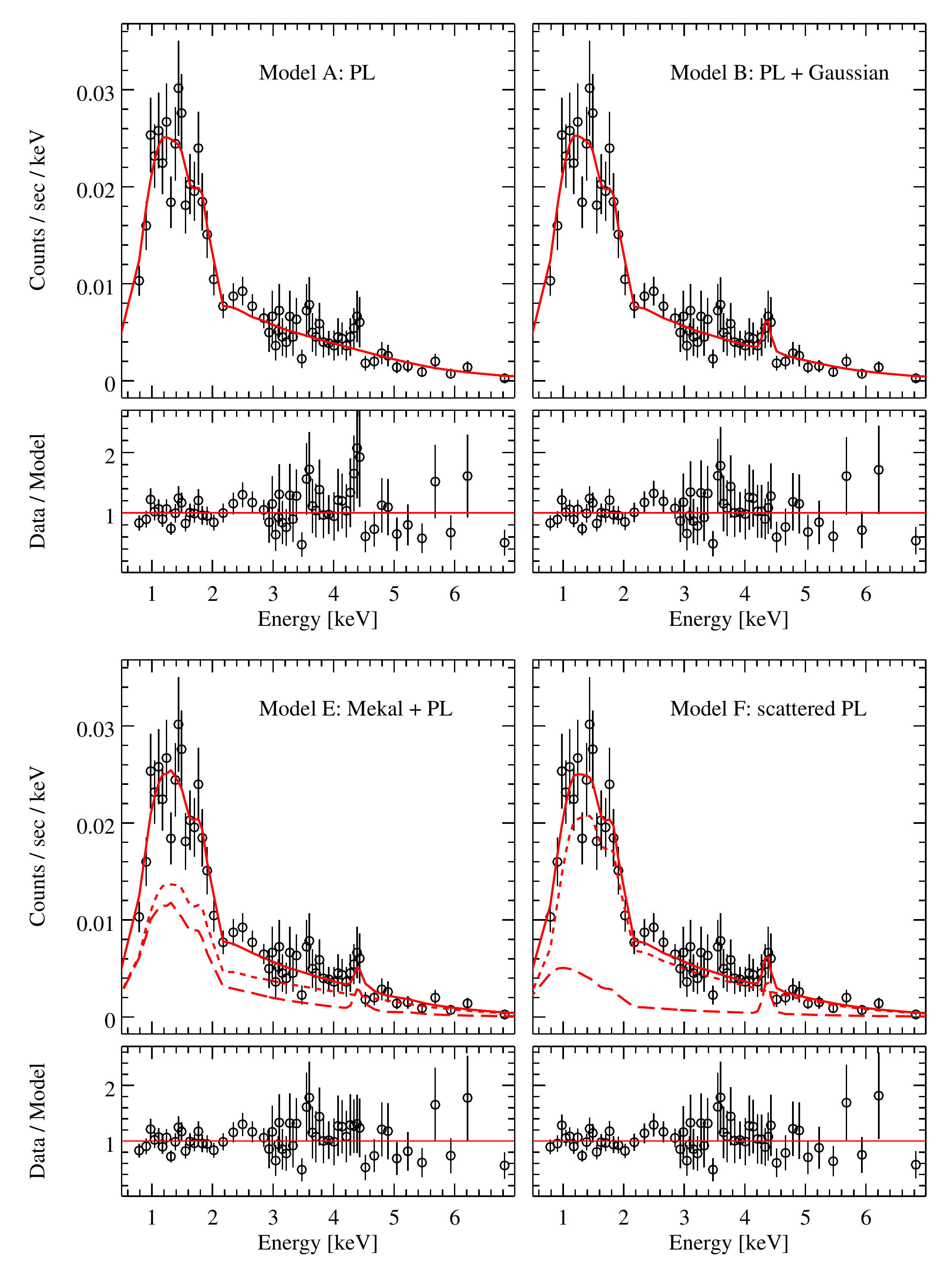}
\caption{Spectral modeling of 0710$+$0439. Each panel shows the {\it Chandra} data (open circles), total models
(solid; see Table~\ref{tab:0710}), model components, and the ratio of the data to the best fitting model.
Top: an absorbed power law without the emission line (left, Model A), and with the emission line (right, Model B).
Bottom: two-component models consisting of (left, Model E) thermal emission of diffuse plasma (long dash) and
a direct power law (short dash); (right, Model F) a direct power law (short dash), and a scattered power
law plus a Gaussian emission line (long dash). The data have been grouped for presentation purposes
by requiring $\rm S/N = 5$ and $\rm S/N = 3$, below and above 3\,keV, respectively.}
\label{fig:0710}
\end{figure*}
\end{center}

\section{Results}
\label{sec:results}

We detected X-ray emission associated with all the CSOs in our {\it
  Chandra} sample. However, because the majority of our {\it Chandra}
observations were short, designed for X-ray detections, the detailed
analysis were only possible for about half of the sources in the
sample.  In the short $\sim$5\,ksec observations the detected source
counts ranged from 3 to 231 counts (see Table~\ref{tab:chandra}).  The
background in such a short {\it Chandra} observation of a point source
is very low and even 3 counts in 0116$+$319 constitute a detection at
high significance level (p-$\rm value < 0.01$).  However, we can only
provide a limiting flux in this case (see Table~\ref{tab:chandra-fit}),
while for the other sources we were able to extract {\it Chandra}
spectrum and apply parametric models.

The three longer exposures provided data of high enough
sensitivity to explore the CSOs X-ray environment on arcsecond scales.
We extracted the X-ray surface brightness profiles assuming
the annuli regions centered on the radio source position to quantify
the presence of an X-ray emission outside the central point source.
In two cases, 0710$+$0439 and 1934$-$638, the profiles are
consistent with the background level outside a $\sim$6$\arcsec$
radius.  The residuals in the innermost circular regions
($r<3\arcsec$), slightly deviate from zero suggesting the possibility of
an additional emission component. However, the formal significance test 
gives a p-value of 0.33, which does not allow us to reject the null
hypothesis that the source is point-like.  We concluded that in these
two sources the observed X-ray emission originates in an unresolved
region within $r <1.5\arcsec$ (corresponding to $< 9.5$\,kpc and
$<4.7$\,kpc for each source, respectively). There is no extended X-ray
emission on large scales detected above the background level of
0.0048\,cts~arcsec$^{-2}$.

The surface brightness profile for the third object, 1718$-$649, is
shown in Figure~\ref{fig:profile}.  An excess of the observed emission
above the PSF profile is visible at the distances within 2--6\arcsec\,
($0.6-1.8$\,kpc) from the radio source position indicating an extended
X-ray emission present in this source. We add another model component
(the standard beta model, {\tt beta1d} in {\it Sherpa}; King 1962) to
account for this emission.  We obtained a 1$\sigma$ lower limit of
$\beta > 0.55$ in our best fit model. Deep \chandra\ observations are
needed to better constrain the properties of this emission.

In general, our baseline absorbed power law model provided a good
description of the X-ray continua of the CSO sources.  The modeling
results are given in Table~\ref{tab:chandra-fit} which 
displays the Galactic and intrinsic absorption columns, photon index,
and the unabsorbed flux in the soft (0.5--2\,keV) and hard
(2--10\,keV; extrapolated from the model fitted over the 0.5--7\,keV band)
X-ray bands. The uncertainties are given as 1$\sigma$ for
one interesting parameter. In seven CSOs the photon index was a free
parameter and we obtained $\Gamma \simeq 1.4$--1.7,  with the exception
of 2021$+$614 and 1511$+$0518 where $\Gamma \simeq 1$.
The intrinsic absorption
component was required at $>2\sigma$ confidence level in three CSOs
(0710$+$439, 1946$+$708, and 0035$+$227; note though that in
0035$+$227 the photon index was fixed at 1.7) with $N_H^z \simeq
(0.6$--$1.7) \times 10^{22}$\,cm$^{-2}$. In three other cases
we were able to derive only upper limits on
the intrinsic column densities (but see Section~\ref{sec:2021}). The
X-ray spectra of the remaining five sources are suggestive of the
presence of an intrinsic absorbing neutral hydrogen column with $N_H^z
\simeq 10^{21}$--10$^{22}$\,cm$^{-2}$, although at low statistical
significance.

Even though we detected an extended X-ray emission in 1718$-$649, the
lowest redshift CSO in our sample,  we find that the X-ray
spectrum of this source is quite well modeled with an absorbed power
law. However, our {\it Chandra} spectrum has only 224$\pm15$ counts
(see Table~\ref{tab:chandra-fit}) and any detailed analysis of
potential contributions from emission associated with the extended
medium would require higher quality X-ray data.

Below we provide a detailed description of the X-ray results for three
particular CSOs in our sample: 0710$+$439 has a hard band narrow
emission line (Section~\ref{sec:0710}); 2021$+$614 and 1511$+$0518 have an extremely
hard X-ray photon index compared to typical AGN
(Section~\ref{sec:2021}); and 1607$+$26 is accompanied by a secondary
X-ray source resolved within the prior {\it XMM-Newton} extraction
region (Section~\ref{sec:1607}). We also comment on 1934$-$63 and
1946$+$708 observed previously with Beppo-SAX and revisited with our
{\it Chandra} program (Section~\ref{sec:beppo}).

\begin{table*}
{\scriptsize
\noindent
\caption[]{\label{tab:0710} Spectral models for 0710$+$439}
\begin{center}
\begin{tabular}{llccccccccc}
\hline\hline
\\
\# & Model           & $\rm N_H(z)$       & $\Gamma$ & kT    & $f_{\rm sc}$ & $\rm E_{Fe\,XXV}$  & EW   & Flux    & Flux  & Cstat/d.o.f.\\
   & description$^a$ & [$10^{22}$\,cm$^{-2}$] &      & [keV] &             & [keV]             & [eV] &  [soft]  &  [hard]  &  \\
\\
\hline
\\
A & PL           & 0.58$\pm0.08$          & 1.59$\pm0.07$          & $\dots$   & $\dots$   &  $\dots$      &   $\dots$  &   $19.7^{+1.5}_{-1.5}$   &   $42.2^{+6.3}_{-5.5}$    & 490.3/443 \\   
B & PL and G     & 0.61$^{+0.04}_{-0.08}$ & 1.64$^{+0.03}_{-0.07}$ & $\dots$   & $\dots$   &  6.62$\pm0.04$  & 154$^{+65}_{-58}$ & $20.2^{+1.6}_{-1.6}$   &   $41.0^{+6.3}_{-5.5}$   & 480.3/441 \\   
C & TB and G     & 0.49$^{+0.07}_{-0.06}$ & $\dots$    &  13.6$^{+2.6}_{-2.8}$ & $\dots$   &  6.63$\pm0.04$  & 164$^{+64}_{-58}$               & $18.0^{+0.7}_{-0.7}$   &   $37.0^{+2.2}_{-3.2}$   & 475.8/441 \\   
\\
\hline
\\
D & PL and Apec  & 0.58$\pm0.08$          & 1.42$^{+0.14}_{-0.20}$ &    5.4$\pm1.5$ & $\dots$   &  $\dots$  &  $\dots$   & $19.7^{+2.9}_{-2.8}$   &   $40.1^{+5.6}_{-5.3}$   &     485.0/441 \\   
E & PL and Mekal & 0.56$\pm0.08$          & 1.39$^{+0.17}_{-0.23}$ &    5.2$^{+2.2}_{-1.3}$ & $\dots$   & $\dots$   &  $\dots$ & $19.7^{+2.9}_{-2.9}$   &   $40.4^{+5.7}_{-5.5}$ & 484.4/441 \\   
\\
\hline
\\
F$^b$ & scattered PL and G & 1.02$^{+0.29}_{-0.22}$ & 1.75$^{+0.11}_{-0.10}$ & $\dots$   & 0.13$\pm0.06$ & 6.62$\pm0.04$ & 1.40$^{+0.97}_{-0.79}\times10^3$ & $23.3^{+3.4}_{-3.3}$   &   $39.8^{+9.8}_{-8.1}$   & 477.2/440 \\   
\\
\hline\hline
\end{tabular}
\end{center}

\smallskip Notes:\\
$^a$ PL -- power law; G -- gaussian line; TB -- thermal bremsstrahlung;
$f_{\rm sc}$ is the fraction of the intrinsic power law scattered into
our line of sight by a photoionized medium that is also responsible for the Fe XXV iron line emission; Unabsorbed flux in 
the soft (0.5--2\,keV) and  hard (2--10\,keV) bands extrapolated
from the model fitted over the 0.5--7\,keV range, in units of [$10^{-14}$\,erg\,cm$^{-2}$\,s$^{-1}$]. \\
A -- {\tt phabs * zphabs * powerlaw}\\
B -- {\tt phabs * zphabs * (powerlaw + zgauss)}\\
C -- {\tt phabs * zphabs * (zbremss + zgauss)}\\
D -- {\tt phabs * zphabs * (apec + powerlaw)}\\
E -- {\tt phabs * zphabs * (mekal + powerlaw)}\\
F -- {\tt phabs * (zphabs * powerlaw +  fsc * powerlaw + zgauss)}\\
All models modified by Galactic absorption with $N_{H} = 8\times 10^{20}$\,cm$^{-2}$;
abundances in {\tt apec} and {\tt mekal} models fixed at 0.5 Solar value;
{\tt mekal} density fixed at 1\,cm$^{-3}$; width of the
Fe line fixed at $\sigma = 0.01$\,keV.\\
$^b$ EW calculated with respect to the unabsorbed scattered power law, {\tt fsc * powerlaw}.
}
\end{table*}


\subsection{0710$+$439}
\label{sec:0710}

The photon index and the amount of intrinsic absorption in our {\it
Chandra} spectrum of 0710$+$439 agree very well with the earlier {\it
XMM-Newton} results of Vink et al. (2006). The source did not vary
between the {\it XMM-Newton} and \chandra\ observations showing
consistent 2--10\,keV flux of  about
$4.1\times10^{-13}$\,erg~cm$^{-2}$~s$^{-1}$. However, we find clear
residuals in the hard X-ray band suggestive of the presence of an
emission line (Figure~\ref{fig:0710}, top/left). To account for this
feature we added a Gaussian component to the absorbed power-law
spectral model (Model B in Table~\ref{tab:0710};
Figure~\ref{fig:0710}, top/right). The line is unresolved ($\sigma =
10$\,eV; fixed) with the best fit line energy of $E = 6.62\pm
0.04$\,keV (rest frame) and the equivalent width $\rm EW =
154^{+65}_{-58}$\,eV. Visual inspection of the {\it XMM-Newton}
modeling presented by Vink et al. (2006) tentatively hints at the
presence of an emission line at similar energy also in the {\it
XMM-Newton} dataset.

Emission lines at the energies within $\sim$6.4--6.9\,keV due to iron
are common in X-ray spectra of accreting black holes
\citep{krolik1987,george1991,zycki1994}, including radio-quiet and
non-blazar radio-loud AGN (e.g., \citealt{hardcastle2009}, Winter et al.
2010; Fukazawa et al. 2011; Zhou, Zhao \& Soria 2011) and Galactic
black hole binaries (e.g., Done et al. 2007).  In particular, a narrow
fluorescent neutral Fe K$\alpha$ line at 6.4\,keV (rest frame) seems
to be omnipresent in the spectra of radiatively efficient AGN (Bianchi
et al. 2004; Nandra et al. 2007; Iwasawa et al. 2012; and references
therein), and is believed to originate from a cold matter, e.g. a
molecular torus, illuminated by hard X-rays.

Narrow $\sim$6.6--6.9\,keV emission lines from an ionized iron have been
observed e.g. in luminous PG quasars, including the radio-loud source
Mrk 1383 (Porquet et al. 2004), and in a substantial fraction of
Seyfert 1 galaxies (Patrick et al. 2012). These lines could be emitted
by a hot ionized region of an accretion disk illuminated by an
external X-ray source \citep[e.g.,][]{matt1993,rozanska2002};
a distant gas,
photoionized by the nuclear illumination (e.g., Bianchi \& Matt 2002);
or a diffuse thermal plasma at $kT \gtrsim 5$\,keV \citep{smith2001}

The best fit parameters of the line detected in 0710$+$439 assuming
Model B (see Table~\ref{tab:0710}) are consistent with the first
scenario and with the He-like Fe XXV line emission\footnote{The line
  is most probably a blend of not resolved resonance (6.700\,keV),
  intercombination (6.682\,keV and 6.668\,keV), and forbidden
  (6.637\,keV) lines.} in a system with an inclination angle $\cos i
\simeq 0.5$--0.6 and an accretion rate $\dot{M} \simeq
0.4$\,$\dot{M}_{\rm Edd}$ (Matt et al.  1993).  However, in order to
study the origin of this emission line in more details we also tested
four other spectral models of increased complexity that are typically
considered for AGN X-ray emission(Table~\ref{tab:0710}). Given the
quality of the current data all the models are statistically
equivalent, but considering the range of the best fit parameters we
can potentially rule out some of them.

First, we parametrized the X-ray continuum with a thermal
bremsstrahlung model ({\tt zbremss})  and kept the Gaussian profile to fit the emission
feature (Model C in Table~\ref{tab:0710}).  We found a statistically
good fit, but with a relatively high temperature, $kT \sim 14$\,keV.
Next, we replaced the bremsstrahlung continuum with a single
temperature plasma ({\tt mekal}) model (metal abundances fixed at
0.5 the Solar value) and found that at this high temperature the
strength of the ionized Fe lines was overpredicted, especially at the
energy of the fully ionized Fe XXVI, while the data did not provide
evidence for the presence of the Fe XXVI line. Similar result was
obtained when we used the {\tt apec} plasma model instead of the {\tt
  mekal} model. Hence we concluded that single component thermal
models are not likely given the physical properties required by the
best fit parameters. Therefore, we tried a two-component model,
consisting of a power law and either {\tt mekal} or {\tt apec}
(Models D and E in Table~\ref{tab:0710}).  We obtained the best fit
plasma temperature $kT \sim 5$\,keV, and a hard photon index, $\Gamma
= 1.4$--1.5, which are consistent with the prediction of jet/lobe
emission.

Finally, we tested a model in which the spectrum is composed of the
intrinsically absorbed primary emission (intrinsic emission modeled
with a power law) and an unabsorbed fraction of the same emission
scattered by a warm medium.  The warm medium is also producing the
iron emission line (Model F in Table~\ref{tab:0710}).  In this scenario
the photon index is $\Gamma = 1.75^{+0.11}_{-0.10}$, the primary
continuum is absorbed with a column of $\rm N_H^z \simeq
10^{22}$\,cm$^{-2}$, and the scattering fraction is $\sim$13\,\%.  The
EW of the line (computed with respect to the scattered continuum) is
$\sim$1.4\,keV, in agreement with the values found by Bianchi \& Matt
(2002) for column densities of the order of a few $\times
10^{21}$\,cm$^{-2}$ (see their Figure~4).  The best fit parameters of
the models A, B, E and F, and the ratio of the data to the models are
presented in Figure~\ref{fig:0710}. We conclude that with the present
data we are not able to state if the soft X-rays and the He-like iron
line in 0710$+$439 are due to scattered nuclear emission (Models B or F) or
thermal emission of the ISM heated through interactions with the 
expanding jet (Models D and E). 

The neutral iron K$\alpha$ line has been previously reported in CSO
sources observed with {\it XMM-Newton} (OQ\,208, Guainazzi et al.
2004; and 1607$+$26, \citealt{teng+2009}), while Risaliti et al.
(2003) detected neutral and highly ionized ($E \simeq 6.9$\,keV) iron
emission lines in 1934$-$638 and 1946$+$708 (respectively) using {\it
  Beppo-SAX}.  These two emission lines are not present in the low
counts {\it Chandra} spectra\footnote{p-values of 0.9 for 1934$-$63
  and 0.4 for 1946$ +$708 when testing for the line with the null
  hypothesis of the power law model, and using the simulations as
  described in \cite{protassov2002} ({\tt plot\_pvalue} in {\it
    Sherpa}).} of 1607$+$26 (see Sec.~\ref{sec:1607}) and of the two
{\it Beppo-SAX} sources (see Sec.~\ref{sec:beppo}).

\cite{muller2015} detected a narrow emission line in a CSO candidate
(PMN\,J1603$-$4904), and derived the redshift of the source assuming
that the most likely rest frame energy of the line is 6.4\,keV. The
presence of a mildly ionized Fe line in the spectrum of 0710$+$439
indicates that there might be a systematic error in the redshift
derivation of PMN\,J1603$-$4904 because of a range of ionization
states of iron producing the emission lines in the CSO sources.

Lastly, we note that 0710$+$439 has the highest number of counts
(1,679) among the sources in our sample. The remaining CSOs were
detected with 3--362 counts. A higher number of counts in {\it
  Chandra} spectra are needed for detecting narrow spectral features.

\subsection{2021$+$614 and 1511$+$0518}
\label{sec:2021}

The absorbed power-law model applied to two CSOs,
  2021$+$614 and 1511$+$0518, resulted in an unusually hard photon
  index of $\Gamma = 0.8\pm0.3$ and $\Gamma = 1.0\pm0.2$
  respectively.  Typically, such a low value indicates a strong
nuclear obscuration able to cover completely the source of the primary
X-ray radiation.  The observed spectrum is then dominated by a Compton
reflection hump (and an accompanying neutral iron line) from a distant
material, presumably a molecular torus, resulting in an artificially
low photon index. AGN with these properties are called `Compton
thick', and the amount of the hydrogen absorbing column required to
distort the X-ray spectrum in this manner is $N_H \gtrsim 10^{24}$
\citep[e.g.,][]{comastri2004}.  The first radio-loud Compton thick AGN
was found in OQ\,208, a CSO source observed with {\it XMM-Newton} by
\cite{guainazzi2004}.  \cite{teng+2009} speculated that 1607$+$26 is
also a Compton thick CSO. They based their argument on the photon
index $\Gamma=0.4\pm0.3$ obtained when an absorbed power-law model was
applied to the {\it XMM-Newton} data of this source. By analogy with
OQ\,208 and 1607$+$26, it is likely that 2021$+$614 and
  1511$+$0518 are Compton thick CSO candidates.  Our result on the
intrinsic absorption in these sources would then be significantly
underestimated because in our fit we do not account for the column
density that would be needed to absorb the primary continuum. Indeed,
when we fit the data of 2021$+$614 with the sum of an intrinsically
absorbed power law and unabsorbed reflection ({\tt pexrav}) component
we obtain {\rm Cstat/d.o.f. = 209/442}, $N_H^z \gtrsim 9.5 \times
10^{23}$\,cm$^{-2}$.  The resulting photon index of the underlying
power law would be, however, very soft, $\Gamma = 3.3\pm0.3$, and the
unabsorbed intrinsic 2--10\,keV luminosity would increase by an order
of magnitude making 2021$+$614 the second CSO in our sample (in
addition to 0710$+$439) to exceed $L_X \sim 10^{44}$\,erg\,s$^{-1}$
(comparable to the Eddington luminosity of a 10$^6$\,M$_{\odot}$ black
hole).

In 1511$+$0518 the Compton thick model results in Cstat/d.o.f. =
198.2/443, $N_H^z = 3.8^{+4.0}_{-1.3} \times 10^{23}$\,cm$^{-2}$, $\Gamma
= 3.8^{+0.3}_{-0.4}$ and $L_X \sim 3 \times 10^{43}$\,erg\,s$^{-1}$.

On the other hand, we do not see any evidence for the presence of a
neutral iron line in the spectra of 2021$+$614  and 1511$+$0518.
  The line is unconstrained in 1511$+$0518.  In 2021$+$614 we can
  only derive an upper limit on the EW of this line, $\rm EW <
0.7$\,keV, inconsistent with  the expected range derived by Matt
et al.  (1996; $\rm EW = 1.3$--2.7\,keV).
Thus, it is interesting to note that the low X-ray photon index and
the low intrinsic absorption can be understood in the context of a
non-thermal X-ray lobe emission model for the GPS/CSO sources
\citep{stawarz2008,ostorero2010}.

Unfortunately, the quality of our present {\it Chandra} data is not
sufficient  to discriminate between the Compton thick absorption
  and X-ray lobe emission scenarios for X-ray emission in 2021$+$614
  and 1511$+$0518 (only 54 and 49 counts collected in 5\,ksec and
  2\,ksec \chandra\ observations, respectively). Deeper X-ray
observations with {\it Chandra} or {\it XMM-Newton} accompanied with
the hard $>10$\,keV band exposures ({\it NuSTAR}) would be needed in
order to resolve this issue.

\subsection{1607$+$26}
\label{sec:1607}

Owing to the superior spatial resolution of {\it Chandra} we resolve
the source region of 1607$+$26 observed with {\it XMM-Newton} by
\cite{teng+2009} into two point-like X-ray sources (see
Figure~\ref{fig:images}). The location of 1607$+$26 matches the
position of the upper left source visible in Figure~\ref{fig:images}
with the detected 213 counts.  The secondary source separated by
$\sim$13.5$\arcsec$ (lower right corner in Figure~\ref{fig:images})
has 32 counts and its spectrum could be modeled with an absorbed power
law (assuming only Galactic absorption since the redshift to the
source is unknown) with a photon index $\rm \Gamma = 1.4\pm0.3$ (see
Table~\ref{tab:chandra-fit}).  This means that the {\it XMM-Newton}
spectrum obtained by \cite{teng+2009} represents a combination of
emission from 1607$+$26 and from the secondary source.  We detect an
order of magnitude decrease between the hard X-ray fluxes of {\it
XMM-Newton} ($4.2 \times 10^{-13}$\,erg\,s$^{-1}$\,cm$^{-2}$) and {\it
Chandra} ($4.8 \times 10^{-14}$\,erg\,s$^{-1}$\,cm$^{-2}$). This
change is larger than the sum of the fluxes from the two sources, and
it is larger than the absolute calibration uncertainties between the
{\it Chandra} and {\it XMM-Newton} satellites
\citep{tsujimoto2011}. Therefore, we concluded that the X-ray flux of
one or both of the sources decreased during the $\sim3$\,years
interval between the two observations.

We searched the SDSS, 2MASS and FIRST catalogs to check the
identification of the secondary X-ray source, but did not find any
corresponding emission. If the source were associated with 1607$+$26
then it would be located $\sim$80\,kpc away along the axis of the CSO
radio structure, on the extension of its radio jet (see
\cite{nagai+2006} for a discussion of the CSO radio morphology). 
It is interesting to speculate that the observed secondary X-ray
source might be a hot spot due to the past radio activity in the
framework of the intermittent scenario for the GPS sources.  However,
assuming the expansion velocity of $0.34c$ (Table~\ref{tab:cso}) the
age of this feature would be $\sim$1\,Myr, while the terminal hot
spots are expected to die very quickly, on timescales $\sim 0.1$\,Myr
once the jet switches off \citep{carilli1988}.  Therefore, the nature
and the association of this secondary source with 1607$+$26 remains
unclear.

Contrary to \cite{teng+2009} who reported $\Gamma = 0.4\pm0.3$
in 1607$+$26, we find $\Gamma=1.4\pm0.1$,
an X-ray photon index that is fairly common in a population of nearby
radio-quiet and radio-loud AGN (e.g. Winter et al. 2009, Fukazawa et
al. 2011).  We do not detect any evidence of the neutral iron
K$\alpha$ line in the spectrum. This line was, however, required by
the {\it XMM-Newton} data of \cite{teng+2009} as part of their Compton
reflection dominated model for this source.

One explanation of this result could be that 1607$+$26 is a
transitional Compton thick/thin AGN. Transitions between the Compton
thick and thin regimes have been previously reported in the so-called
`changing look' AGN (see, e.g., \cite{matt2003};
Risaliti et al. 2005; 2010; 2011; and references therein).
The Compton thick/thin switches are thought to be due to an X-ray
absorber made of clouds with $N_H \simeq
10^{23}$--10$^{24}$\,cm$^{-2}$ crossing our line of sight on
timescales between hours and weeks.  

If the flux from the secondary source were constant between the two
exposures, then 1607$+$26 would have been $\gtrsim 10$ times brighter
in the reflection dominated state recorded with {\it XMM-Newton} than
in our observation. This clearly contradicts the `changing look' AGN
scenario. To reconcile the hard X-ray flux variability with the
'changing look' hypothesis, 1607$+$26 could not have contributed more
than $\sim$200\,cts to the {\it XMM-Newton} data, meaning that, the
secondary X-ray source would have to be $\gtrsim 10$ times brighter
than 1607$+$26 in the {\it XMM-Newton} observation. This would imply
that the X-rays detected and modeled by Tengstrand et al. (2009;
including the iron line) originated from the secondary source rather
than from 1607$+$26, eliminating any spectral information about the
state of 1607$+$26 during the {\it XMM-Newton} observation. It would
also indicate that the photon index of the secondary source changed
from $\Gamma\sim0.8$ to $\sim1.4$, its iron line disappeared, and the
hard flux increased by a factor of $\gtrsim 70$ between the two
exposures. These spectral properties would be very difficult to
explain, especially with no radio, optical and $\gamma$-ray
identification available at present for the secondary source. Thus, we
reject the possibility that X-ray variability observed in 1607$+$26 is
due to the Compton thick/thin switches.

The alternative explanations include fading of the nuclear and
reflected 1607$+$26 emission, or a heavy obscuration with a column
density $\gtrsim 10^{24}$\,cm$^{-2}$, as a result of which {\it
Chandra} would observed thermal emission of the ISM. Finally, in the
non-thermal scenario of \cite{stawarz2008} and \cite{ostorero2010}
any potential intrinsic variability of the X-ray continuum
is expected to take place on timescales much too long to account for
the results of the {\it Chandra} observations of 1607$+$26.

Clearly, future X-ray monitoring of this young radio source is needed
in order to determine its nature and X-ray variability properties.

\subsection{1934$-$63 and 1946$ +$708}
\label{sec:beppo}

\citet{risaliti2003} reported that 1934$-$63 and 1946$+$708
  showed {\it Beppo-SAX} spectra typical of the Compton-thick sources
  with $N_H> 2.5\times 10^{24}$\,cm$^{-2}$, strong Fe K$\alpha$ line,
  and the reflection hump. They also considered Compton-thin absorbers
  as a secondary explanation of the data.  We have revisited both
  sources with {\it Chandra}. We could only obtain a 3$\sigma$ upper
  limit on the line equivalent width of $<0.96$\,keV in 1934$-$63 and
  of $<7.2$\,keV in 1946$ +$708, assuming that the lines are
  unresolved ($\sigma = 10$\,eV) and located at the rest energies,
  6.3~keV and 6.9~keV, respectively (as in \citealt{risaliti2003}).
  Future more sensitive X-ray observations are needed to constrain the
  parameters of these lines.

  We measure a relatively low intrinsic absorption column density in
  both sources ($N_H < 2\times 10^{22}$\,cm$^{-2}$) and do not
  confirm their Compton-thick nature (see Table~\ref{tab:chandra-fit}). We
  conclude that either the {\it BeppoSAX} spectra have been
  contaminated by other X-ray sources in the field, given the 2~arcmin
  source extraction regions in the {\it BeppoSAX} observations, or the
  Compton-thin model is more appropriate explanation of the data,
  or that these sources belong to the AGN class that switches between
  the Compton thick and thin states.

\section{Discussion}
\label{sec:discussion}

We have investigated properties of the X-ray emission 
in a sample of 16 young radio sources with known redshift and
kinematic age measurements of their radio structures.  All sources in
our sample have symmetric radio morphology, have been classified as
CSOs and have been associated with radio galaxies. They do not display
strong AGN emission in the optical band because they either do not
contain an optical-UV bump characteristic of the AGN, or because the
optical emission has been absorbed by a dense medium in the central
region of the host galaxy.  We detected six of the sources in our
sample for the first time in X-rays.

\subsection{Models for CSO X-ray emission}
\label{sec:x-ray}

The angular sizes of the CSO radio emission are of the order of
milliarcseconds \citep[e.g.][]{readhead1996b},
and as such are too small to be resolved with {\it Chandra}.  As a
result, X-ray emission associated with any compact radio jets, lobes,
or bow shocks potentially present around the lobes is integrated
within the 1.5$\arcsec$ radius of the circular source extraction
region, corresponding to 0.4--11\,kpc in our sources, depending on
redshift.  Consequently, the CSOs X-ray emission contains information
on both the circumnuclear region (on sub-pc and pc scale), and the
surrounding ISM (kpc scale).
Different physical processes may, thus, contribute to the observed
X-ray emission \citep[see, e.g.][for a review]{siemiginowska2009}. The
CSOs X-ray emission could arise from the AGN nucleus (X-ray corona)
indicating the nature of the nucleus and its accretion state
\citep{guainazzi2004,vink+2006,siemiginowska+2008,teng+2009}.
However, if the absorption column is larger than $\sim$
10$^{24}$\,cm$^{-2}$, the nucleus could be hidden and the observed
X-rays would be dominated by the thermal emission of the hot ISM from
shocks driven by the expanding jet \citep{heinz1998,odea2000}, or by a
reprocessed nuclear emission due to the reflection and/or scattering
effects \citep[e.g.][]{guainazzi+2006}. On the other hand X-rays can
also be emitted by compact radio lobes through inverse-Comptonization
of local radiation fields
\citep{stawarz2008,ostorero2010,migliori+2012}.  These models result
in different predictions on the X-ray photon index, amount of
intrinsic X-ray absorption, and the luminosity and could be tested
with deep observations.

The CSOs in our sample are rapidly expanding, as demonstrated by their
measured hot spot expansion velocities spanning the 0.07--0.4\,$c$
range (Table~\ref{tab:cso}, and references therein). Based on these
velocities, the CSOs kinematic age estimates range between $\sim$100
and 3000\,yrs.  These are, thus, the youngest radio sources which can
provide information about the accretion state of supermassive black
holes shortly after the onset of the jet formation, as well as about
the properties of the environment into which the newly born jets are
evolving, and with which they are interacting.

\subsection{Environment of CSO}
\label{sec:environment}

Jets and lobes in young radio sources, confined to the central parts
($\sim 1$\,kpc) of their host galaxies, should interact with a denser
and less homogeneous medium than their evolved ($10s-100s$\,kpc-long)
analogs.  Indeed, our results support this idea. We detected
multi-phase galactic environments consisting of diverse gaseous
components co-spatial with the young expanding CSO radio structures:
(1) the extended thermal component in 1718$-$649 (Section~\ref{sec:images},
Figure~\ref{fig:profile}); (2) possibly an ionized diffuse plasma giving
rise to the He-like Fe emission line in 0710$+$439
(Section~\ref{sec:0710}, Figure~\ref{fig:0710}); (3) neutral hydrogen with
moderate column density, $N_H^z \simeq 10^{21}$--10$^{22}$\,cm$^{-2}$,
in the majority of the sources (Table~\ref{tab:chandra-fit},
Figure~\ref{fig:ages}, middle); (4) possibly a dense Compton-thick
absorber  in 2021$+$614  and 1511$+$0518
(Section~\ref{sec:2021}).


\begin{figure}
\includegraphics[width=\columnwidth]{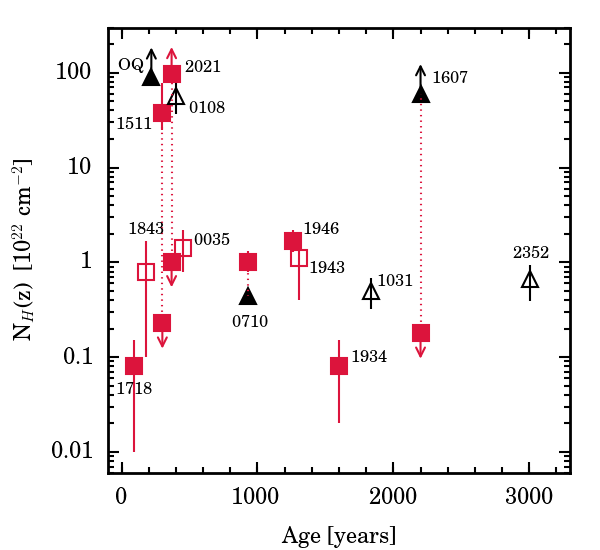}
\includegraphics[width=\columnwidth]{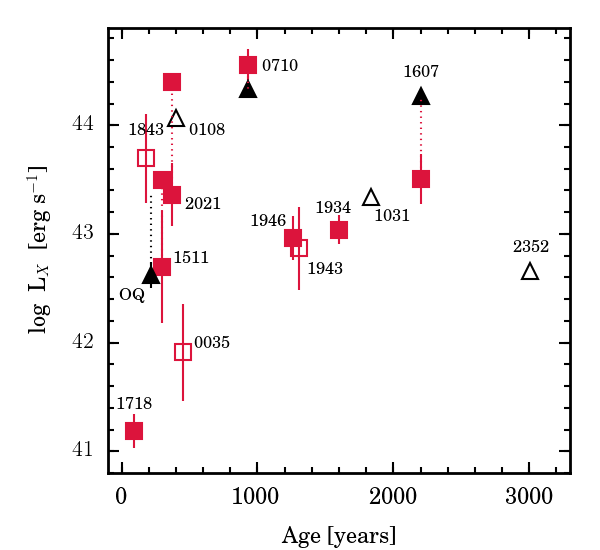}

\caption{ {\bf Top:} Intrinsic X-ray absorbing column vs.
  age based on our {\it Chandra} results (squares) and {\it XMM-Newton}
  results (triangles; sources 13--16 in Table~\ref{tab:cso};
  filled and open symbols indicate fits with $\Gamma$ free and fixed,
  respectively; vertical lines connect multiple measurements (or
  measurements obtained through alternative models as in 
  Section~\ref{sec:2021}) available for the same source. {\bf Bottom:}
  Intrinsic X-ray luminosity (2--10\,keV) vs. age based on our {\it Chandra}
  results (squares, K-corrected) and {\it XMM-Newton} results (triangles; sources
  6, 10, 13--16 in Table~\ref{tab:cso}). Symbols as in the top panel;
  note the large scatter within the youngest sources. }
\label{fig:agedependence}
\end{figure}

Recent multi-wavelength studies also provide evidence that CSO radio
sources are embedded in complex gaseous environments. In the IR band
the Spitzer data show a variety of the IR line properties indicating
an enhanced star formation rates in host galaxies of compact radio
sources \citep{willett2010, guillard2012, dicken2012}.  One
explanation could be a jet-induced star formation in CSOs
\citep{labiano2008}, suggestive of the interactions between the jet
and dense environment. \cite{tadhunter2011} points to a possible
selection bias in flux limited radio samples that impacts general
conclusions about the epoch of star formation and triggering the radio
source. However, these IR results are consistent with the CSO source
being embedded in a dense environment.

In the radio band, \cite{gereb2014,gereb2015} show that the properties
of the HI absorption line in the spectra of GPS and compact steep
spectrum sources (CSS; a more evolved stage compared to the GPS
sources) differ substantially from those of the extended radio
sources.  Based on the correlation between the amount of HI absorption
and the type of a radio source ($\sim$55\% HI detection rate, and
optical depths higher in the compact sources than in the extended
sources) the authors argued that the compact radio sources are
embedded in a medium that is rich in atomic gas. Moreover, the HI
kinematics in the GPS/CSS sources (broader and more asymmetric,
blue-shifted profiles) suggested that the gas is unsettled and may form
outflows. They concluded that the evidence of interactions between the
radio sources and rich ambient environment is more frequent in the
young/compact sources than in the extended sources.
 
Finally, studies of the relationship between the X-ray and radio
absorbing columns reveal a correlation between the $\rm NH$ and
$\rm NHI$ absorption in GPS/CSO sources \citep{ostorero2015},
suggesting that the X-ray and radio absorbers are co-spatial.  This is
an important constraint, given the latest radio results from the low
frequency radio observations \citep{callingham2015}. They seem to rule
out the synchrotron self absorption process and favor the free-free
absorption as the origin of the radio obscuration in the GPS source,
PKS\,B0008$-$42. This again points to a rich, multi-phase
environment impacting the evolution of CSOs.

It is, thus, interesting to interpret our findings on the CSOs
intrinsic neutral hydrogen column density in the context of the
absorption properties of an unbiased population of AGN.  We compare
our results with those obtained by \cite{burlon2011} for their
subsample of Sy1.8--2 galaxies from the hard X-ray Swift/BAT survey
(note that CSOs are observed at large inclinations to the jet axis,
and as such should be compared with type 2 AGN).  In the sample of
CSOs presented in Table~\ref{tab:cso} (see
 Figure~\ref{fig:agedependence} top) we observe that 2 of 15 (13\%)
CSOs show evidence for Compton-thick absorption with $N_H^z \gtrsim
5\times10^{23}$\,cm$^{-2}$ (33\% if we include 2021$+$614, 1511$+$0518
and 1607$+$26 proposed to be Compton-thick candidates in this work and in 
\citealt{teng+2009}), respectively, as compared to $\sim$20\% in Sy1.8-2 galaxies in
\cite{burlon2011}. Conversely, 9 of 15 (60\%) sources have $N_H^z <
10^{22}$\,cm$^{-2}$ (40\% if we exclude 2021$+$614, 1511$+$0518 and 1607$+$26).
Instead, only 7\% of the Sy1.8--2 galaxies have column densities in
this range.  While the fraction of Compton-thick CSOs seems to be in a
relatively good agreement with that found in a Sy1.8--2 population of
\cite{burlon2011}, especially given the small size of our CSOs sample,
the overabundance of sources with only moderate intrinsic obscuration
seems puzzling.  \cite{hardcastle2009} pointed out that the Narrow
Line Radio Galaxies (NLRG), considered to be the more evolved
analogues of CSO, have $N_H^z \gtrsim 3\times 10^{22}$\,cm$^{-2}$
associated with the absorption of the nuclear emission by a molecular
torus. Consequently, the large fraction of CSOs with hydrogen column
densities lower than those typically found in NLRGs, may indicate ---
echoing the conclusions of \cite{ostorero2015} --- that the CSO X-ray
obscuration is caused by the gas located on much larger scales than
molecular tori.

\subsection{X-ray continuum of CSO}

The 2--10\,keV  X-ray luminosities of our sources
cover $>$3 orders of magnitude (Figure~\ref{fig:agedependence} bottom), $L_X
\simeq 2\times10^{41}$--$6\times10^{44}$\,erg\,s$^{-1}$, typical for
the $z<1$ AGN 
\citep[e.g][]{akylas2006,winter2009,fukazawa2011,jia2013,aird2015}.
Interestingly, we observed that the oldest CSOs ($t > 1000$\,years)
tend to be relatively bright in X-rays, with luminosities
$L_X>10^{43}$\,erg\,s$^{-1}$, while the youngest CSOs can be found at
any X-ray luminosity covered by the sample.

We found hard X-ray spectra with a photon index $\Gamma \sim 1.4$--1.7
in five (out of seven) sources, where we were able to
model the photon index.  Among non-blazar AGN, this range of photon indices is often seen in
radio-quiet AGN, although it appears to be the most compatible with
the average photon index of the non-blazar radio-loud AGN
(\citealt{reeves2000,kelly2008,sobolewska2009,sobolewska2011} but see
\citealt{young2009}).
The difference between the average X-ray spectral slopes in radio-loud
and radio-quiet AGN populations may result from either a dilution of
the X-ray coronal emission, or a non-negligible contribution from the
non-thermal emission of jets \citep{belsole2006}. 
It is, thus, likely that in the particular case of CSOs the X-ray jets
and/or lobes contribute to the observed X-ray radiation, as suggested
by \cite{stawarz2008} and \cite{ostorero2010}.
The photon indices we observed would then be consistent with a
non-thermal scenario in which low-energy photons are inverse Compton
upscattered to X-ray energies by electrons with a power-law spectral
energy distribution with a rather standard spectral index, $p\sim2$.

Thus, the shape of the absorbed power law X-ray continuum that we
derived for the CSOs in our sample is consistent with either the
nuclear (X-ray corona) emission absorbed on the torus scale
($\lesssim$\,pc), or an extended (X-ray jets/lobes) emission absorbed
on the host galaxy ($\sim$\,kpc) scale. We note that the results
described in Section~\ref{sec:environment}, along with that of
\cite{ostorero2015}, favor the latter possibility. Instead, high
intrinsic luminosities, exceeding 10$^{44}$\,erg\,s$^{-1}$, derived
for several CSOs (Figure~\ref{fig:agedependence}, bottom panel),
together with the likely detection of the reprocessed/scattered
nuclear light in 2021$+$614, 1511$+$0518 (Section~\ref{sec:2021})
and 0710$+$439 (Section~\ref{sec:0710}) may indicate the dominant AGN
contribution to the observed X-ray emission of young radio sources.

Our results do not appear to favor thermal emission of the diffuse ISM
plasma as the origin of the CSO X-ray continuum, unless the soft
X-rays and the He-like iron emission in 0710$+$439 are interpreted as
a sole example in which the ISM is heated by the expanding radio
structure (Section~\ref{sec:0710}, Figure~\ref{fig:0710}, Model E).

\section{Summary and Conclusions}
\label{sec:conclusions}

We examined the X-ray properties of 16 young radio sources with {\it
  Chandra} and observed variety of interesting characteristics, which
we summarize below.

\begin{itemize}

\item We found that our CSO sample covers a wide range in X-ray
luminosity, $L_{2-10\,\rm keV} \sim 10^{41}$--$10^{45}$\,erg\,s$^{-1}$
and that the majority of sources contains a modest amount of the
intrinsic absorption ($N_H \simeq 10^{21}$--10$^{22}$\,cm$^{-2}$).  

\item  We obtained the highest quality X-ray spectrum of the CSO source to
  date in our long {\it Chandra} observation of 0710$+$439 and found
  an ionized iron emission line, $E_{rest}=(6.62\pm0.04)$\,keV.
  However, we were not able to determine the origin of the line, as it
  could be related to the emission from the hot ionized accretion disk
  illuminated by an external source, from a distant photoionized
  medium, or from a diffuse hot thermal plasma within the nuclear
  region. Depending on the model, we derived EW of the line ranging
  from $154^{+65}_{-58}$\,eV (Model B) to $1.40^{+0.97}_{-0.79}$\,keV
  (Model F).

\item We found a hard photon index of $\Gamma = 0.8^{+0.3}_{-0.2}$ in
  2021$+$614 and $\Gamma = 1.0\pm0.2$ in 1511$+$0518 consistent
  with either a Compton thick absorber or non-thermal emission from
  compact radio lobes.

\item We discovered that the 2--10\,keV X-ray flux decreased by an order
of magnitude since the 2008 {\it XMM-Newton} observation of 1607$+$26.
Future monitoring of this source is needed in order to characterize
and understand the nature of the variability.

\item We examined the highest angular resolution X-ray images of the CSOs and
detected extended X-ray emission in 1718$-$649 which needs to be
studied with the deeper X-ray observations in the future.

\end{itemize}

\medskip

We found that short $\sim$5\,ksec X-ray observations with ${\it
Chandra}$ have proven to be effective in detecting the CSOs sources
and deriving their basic X-ray properties. With our {\it Chandra}
program, we observed six CSOs for the first time in X-rays, and we
increased the size of the known CSOs sample with measured redshift,
kinematic age, and X-ray information by $\sim$68\%.  The CSO X-ray
population can be further extended through snapshot {\it Chandra}
and/or {\it XMM-Newton} observations of seven sources whose kinematic
ages have been recently derived (An \& Baan 2012,
Figure~\ref{fig:ages}).  However, it is apparent that deeper exposures
are needed in order to study details of the CSOs environment (e.g.,
the amount of intrinsic absorption, the gaseous media co-spatial with
the radio sources) and the origin of the CSOs intrinsic X-ray emission
(diffused thermal vs.  nuclear emission, evidence for reprocessing of
the nuclear X-rays by a circumnuclear matter, thermal vs. non-thermal
electron population generating the high-energy CSOs spectra, long term
X-ray variability).  Deep X-ray observations are also necessary in
order to place critical constraints on current theoretical models for
the earliest stage of radio source evolution.

\section*{Acknowledgments}

We thank the anonymous referee for detailed comments.
This research is funded in part by National Aeronautics and Space
Administration contract NAS8-03060.  Partial support for this work was
provided by the NASA grants GO1-12145X, GO4-15099X, NNX10AO60G.  G.M.
acknowledges the financial support from the UnivEarthS Labex program
of Sorbonne Paris Cit\'e (ANR­10­LABX­0023 and
ANR­11­IDEX­0005­02) L.O.  acknowledges support from the INFN
grant INDARK, the grant PRIN 2012 ``Fisica Astroparticellare Teorica''
of the Italian Ministry of University and Research, and the
``Strategic Research Grant: Origin and Detection of Galactic and
Extragalactic Cosmic Rays'' funded by the University of Torino and
Compagnia di San Paolo.  {\L}.S. was supported by Polish NSC grant
DEC-2012/04/A/ST9/00083. This research has made use of data obtained
by the {\it Chandra} X-ray Observatory, and software provided by{\it
Chandra} X-ray Center (CXC) in the application packages CIAO, ChIPS,
and {\it Sherpa}.


\end{document}